# Uncovering vacuum level in infinite solid by real-space potential-unfolding

Duk-Hyun Choe, Damien West, and Shengbai Zhang[*]

*Department of Physics, Applied Physics & Astronomy, Rensselaer Polytechnic Institute,*

*Troy, NY 12180, USA*

*Correspondence to: zhangs9@rpi.edu.

**Abstract**

Although real materials are finite in size, electronic structure theory is built on the assumption of infinitely large solid, which led to a longstanding controversy: where is the vacuum level? Here, we introduce an analytic real-space potential-unfolding approach to uncover the vacuum level in infinitely large solid. First-principles calculations show that, in the absence of a physical surface, the bulk band structure, often measured with respect to an average bulk potential, is offset by a hereto unknown and orientation-dependent bulk quadrupole with respect to the vacuum level. By identifying intrinsic contributions of a bulk solid to its surface and interface properties, our theory eliminates the ambiguities surrounding the physical origin of the band alignment between matters.

Modern electronic structure theory [1,2] assumes infinitely large, periodic systems to take advantage of the Bloch theorem [3] and, for computational physics, the advanced algorithms of fast Fourier transform. The theory has served the condensed-matter community tremendously well [1,2], however, the infinite-crystal assumption leads to many subtleties in electrostatics. One such example is the electric dipole in homogeneous bulk solids [4]. While the dipole is well defined in finite systems, it becomes ill-defined in infinite systems. Nonetheless, it was shown by Resta that the *change* in the dipole under the adiabatic condition, i.e., the bulk polarization, is well-defined as a bulk quantity [5], which was expressed by King-Smith and Vanderbilt [6] in terms of the Berry phase of the wavefunctions [7,8]. Another example is the electric dipole at heterogeneous interfaces, i.e., the built-in potential, $\psi$. Although charge density has been commonly used to investigate $\psi$ [9], there has been a contentious issue regarding the initial basis of comparison [10-14]. Recently, we showed that $\psi$ can be explicitly given by the electrostatic potential. Key in the finding is the identification of a common energy reference $V^0$ between two dissimilar bulk solids as the maximum values of the planar average electrostatic potentials before they were put into contact [15].

Yet, one of the most puzzling problems in electrostatics of infinitely large systems is the average electrostatic potential, $\overline{V}$, with respect to the vacuum. Its ambiguity was first introduced perhaps by Ihm, Zunger, and Cohen in 1979 [16]. After years of debate [17,18], there is now a general consensus that $\overline{V}$ is an ill-defined quantity because the vacuum level in bulk cannot be unambiguously determined. The ambiguity can be alternatively explained in two ways: (1) a direct integration of charge density (see Eq. 2 for $\overline{V}$ below) is a conditionally convergent quantity whose value depends explicitly on how the integral is taken. (2) Equivalently, in the Fourier representation of the electrostatic potential, the limit $\overline{V} = \lim_{\mathbf{G} \to 0} V(\mathbf{G})$ is not generally well defined, but instead depends on the direction from which the limit is taken. In practice, $\overline{V}$ is often set to zero. The advantage is that the eigenvalues with respect to $\overline{V} = 0$, defined here as $\varepsilon_{\overline{V}}$, become orientation independent. However, the alignment between different systems is now ill defined [12,19]. Even worse, calculation shows that $\varepsilon_{\overline{V}}$ is strongly pseudopotential dependent with a deviation that can be as large as several eVs.

Ironically, a number of widely-adopted concepts in condensed-matter physics and chemistry such as deformation potential [19-22], band alignment [23-27], charged defect formation energy [28,29], electrode potentials [13,14,30], and redox potential [30,31] all rely on referencing the energies between different periodic systems. Moreover, understanding the effect of many-body interactions on the electronic states of bulk materials also requires a universal energy reference [32]. Historically, these problems have been dealt with by using non-universal references such as the local (or near) vacuum level of a surface [9,12], core energy level [21], onsite energy (in the tight-binding model), or even the ambiguous $\overline{V}$, which restrict our ability to obtain a complete, or in some cases a correct, answer. Determination of the vacuum level position in infinitely large systems and clarification of the physical meaning of $\overline{V}$ will thus provide fundamental breakthroughs in our understanding of electronic materials.

In this paper, we devise an analytic approach to determine the vacuum level position in infinitely large systems. We find that the average potential with respect to the vacuum level is exactly equal to the quadrupole per volume of the unit cell under consideration. We also show that the eigenvalue of an infinitely large solid can be separated into two parts: the part with respect to $\overline{V}$ that is independent of the crystallographic orientation and the part that is explicitly orientation dependent due to the bulk quadrupole. This procedure paves the way for extracting intrinsic physical properties of any heterogeneous complex systems.

The electrostatic potential at a location in space may be viewed as the least amount of work required to move a positive unit charge from a reference point to a specific point in that space. By the definition, a reference must be provided in order to determine the value of the electrostatic potential. Assuming $4\pi\varepsilon_0 = 1$, the electrostatic potential for a finite system arising from a charge density $\rho(\mathbf{r})$ is given by

$$V(\mathbf{r}) = \int \frac{\rho(\mathbf{r}')}{|\mathbf{r}-\mathbf{r}'|} d\mathbf{r}', \qquad (1)$$

where vacuum is defined as a point infinitely far away from $\rho(\mathbf{r})$, and satisfies $V(|\mathbf{r}|\to\infty) = 0$. For an infinitely large solid, one usually extends Eq. (1) in space so the average electrostatic potential becomes

$$\overline{V} = \frac{1}{\Omega} \int_{\text{unit cell}} V_\infty(\mathbf{r}) d\mathbf{r} = \frac{1}{\Omega} \int_{\text{unit cell}} \int_\infty \frac{\rho(\mathbf{r}')}{|\mathbf{r}-\mathbf{r}'|} d\mathbf{r}' d\mathbf{r}, \qquad (2)$$

where $V_\infty(\mathbf{r})$ is the electrostatic potential due to an infinitely-extended charge density, $\rho(\mathbf{r})$. Although the reference does not change in extending the finite system to an infinite one, the reference becomes buried as there is no longer a region of space where the vacuum is preserved.

Moreover, the evaluation of Eq. (2) yields ambiguities, which may be categorized into four groups: first, there is an infinite number of ways to choose the shape of the unit cell; second, there are an infinite number of ways to choose the origin; third, there are an infinite number of ways to choose the three-dimensional (3D) coordination system for the integration of $\rho(\mathbf{r}')/|\mathbf{r}-\mathbf{r}'|$ in an infinite 3D space; and forth, for a given coordinate system, the order of integration may matter: there are in fact 6 different ways to order the integration along the 3 major axes. Even if we stick to the conventions in electronic structure calculations, namely, a parallelepiped unit cell made of the three lattice vectors and a coordinate system of which the lattice vectors are the axes [1,2], there still exists an infinite number of different types of unit cells to choose from [2].

The ambiguities in Eq. (2) arise because of the difficulty to directly calculate $V_\infty(\mathbf{r})$. Key to eliminate these ambiguities lies in a transformation of Eq. (2), as detailed below. In a system with infinite number of unit cells, $V_\infty(\mathbf{r})$ can be expressed as the superposition of the electrostatic potentials due to the constituent unit cells, i.e., $V_\infty(\mathbf{r}) = \sum_n^\infty V_{n\text{th unit cell}}(\mathbf{r})$, where $V_{n\text{th unit cell}}(\mathbf{r})$ is the potential due to charge densities in the $n$th unit cell (see Fig. 1a). Owing to the translational symmetry, the superposed potential $V_\infty(\mathbf{r})$ in a unit cell (see Fig. 1b) can be unfolded to the potential $V_{\text{unit cell}}(\mathbf{r})$ in the extended space (see Fig. 1c). This procedure is analogous to the folding and unfolding of a band structure, e.g., that of a free-electron gas, in the reduced and extended Brillouin zones in Fourier space [1]. This transforms the difficult integration of $V_\infty(\mathbf{r})$ in a unit cell to an equivalent integration of $V_{\text{unit cell}}(\mathbf{r})$ in an infinite space, namely,

$$\int_{\text{unit cell}} V_\infty(\mathbf{r}) d\mathbf{r} = \int_\infty V_{\text{unit cell}}(\mathbf{r}) d\mathbf{r}. \tag{3}$$

There are two important consequences: first, it is easier to calculate $V_{\text{unit cell}}(\mathbf{r})$ than $V_\infty(\mathbf{r})$, and second, it recovers the vacuum level in real space, when $|\mathbf{r}| \to \infty$, $V_{\text{unit cell}}(|\mathbf{r}| \to \infty) = 0$. Note that this is the vacuum level of the infinitely large system, to be termed the *ideal vacuum level*, $\phi_0$. While by definition, $\phi_0$ should be the true vacuum level in a finite system, it is different from the so-called local vacuum associated with the work function measurement [9,12,33], as the latter includes the effects due to surface electronic and ionic relaxations from those of a truncated bulk.

We consider the evaluation of $\overline{V}$ for the charge density consisting of an arbitrary set of $N$ point charges in a unit cell, $\rho(\mathbf{r}') = \sum_{n=1}^{N} q_n \delta(\mathbf{r}' - \mathbf{r}_n)$, with a corresponding $V_\infty(\mathbf{r}) = \sum_{\mathbf{R}_j} \int_\infty \sum_n q_n \delta(\mathbf{r}' - \mathbf{r}_n - \mathbf{R}_j) / |\mathbf{r} - \mathbf{r}'| d\mathbf{r}'$ and $V_{\text{unit cell}}(\mathbf{r}) = \int_\infty \sum_n q_n \delta(\mathbf{r}' - \mathbf{r}_n) / |\mathbf{r} - \mathbf{r}'| d\mathbf{r}'$, where $\mathbf{R}_j$ is the location of the $j^{\text{th}}$ cell. As $\overline{V} = \Omega^{-1} \int_{\text{unit cell}} V_\infty(\mathbf{r}) d\mathbf{r}$, Eq. (3) yields

$$\overline{V}_{ijk} = \frac{1}{\Omega} \int_{-\infty}^{\infty} \int_{-\infty}^{\infty} \int_{-\infty}^{\infty} V_{\text{unit cell}}(\mathbf{r}) dx_i dx_j dx_k = \frac{1}{\Omega} \int_{-\infty}^{\infty} \int_{-\infty}^{\infty} \int_{-\infty}^{\infty} \sum_{n=1}^{N} \frac{q_n}{|\mathbf{r} - \mathbf{r}_n|} dx_i dx_j dx_k, \quad (4)$$

where the subscript $ijk$ denotes the order of integration to be arbitrary axes $x_i, x_j, x_k$, respectively. As each term diverges, we add and subtract a counter charge at the origin, yielding,

$$\overline{V}_{ijk} = \frac{1}{\Omega} \int_{-\infty}^{\infty} \int_{-\infty}^{\infty} \int_{-\infty}^{\infty} \left[ \sum_{n=1}^{N} \left( \frac{q_n}{|\mathbf{r} - \mathbf{r}_n|} - \frac{q_n}{|\mathbf{r}|} \right) + \sum_{n=1}^{n} \frac{q_n}{|\mathbf{r}|} \right] dx_i dx_j dx_k, \quad (5)$$

where the last term vanishes as the net charge in the unit cell is zero, leaving a sum of finite non-vanishing neutral terms. Using the linearity of integration, Eq. (5) becomes

$$\overline{V}_{ijk} = \frac{1}{\Omega} \sum_{n=1}^{N} \int_{-\infty}^{\infty} \int_{-\infty}^{\infty} \int_{-\infty}^{\infty} \left( \frac{q_n}{|\mathbf{r} - \mathbf{r}_n|} - \frac{q_n}{|\mathbf{r}|} \right) dx_i dx_j dx_k. \quad (6)$$

For an orientation $x_3 = x_1 \times x_2$, where $x_1$ and $x_2$ axes are any of $x_i, x_j, x_k$, Eq. (6) becomes $\overline{V}_{123} = -2\pi Q_{33}/\Omega$, where $Q_{33}$ is an element of the bulk quadrupole tensor, defined by $Q_{\alpha\beta} = \int_{\text{unit cell}} \rho(\mathbf{r}) x_\alpha x_\beta d\mathbf{r}$. More concisely,

$$\overline{V}_\mathbf{n} = -\frac{2\pi}{\Omega} \mathbf{n}^T Q \mathbf{n}, \quad (7)$$

, where $\mathbf{n} = x_3$, which is consistent with the momentum space derivation of $\overline{V}$ (Supplemental Material). The calculation of $\overline{V}$ in absolute terms, i.e., relative to the vacuum level, is thus equivalent to the calculation of $\mathbf{n}^T Q \mathbf{n}$. Here we note that Eq. (7) is equally well applied to continuous charge densities if we let $q_n \to \delta q_n$ and $N \to \infty$. Note that only when the dipole of the unit cell vanishes, the average potential is independent of the choice of origin [18].

The dependence on the order of the last integration in $\overline{V}$ has, in fact, a clear physical meaning, namely, the plane over which the planar average potential is evaluated. For instance, when a $z$-axis integration is performed in Cartesian coordinate system using Eq. (4), $\overline{V}$ is given by $\Omega^{-1} \int_{\text{unit cell}} V(z) dz$, where $V(z)$ is the *xy*-planar average electrostatic potential. On the other hand, in a momentum-space approach, the average potential depends on the direction from which the limit is taken (see Supplemental Material), which along a direction parallel to the *z*-axis can be written as $\lim_{G_z \to 0} V(0,0,G_z) = -2\pi Q_{zz}/\Omega$. The central message here is that, although the vacuum level $\phi_0$ in infinite solid is well-defined, $\bar{V}$ is orientation dependent due to the fact that the often-non-zero quadrupole $\mathbf{n}^T \mathbf{Q} \mathbf{n}$ depends on the orientation.

Note that a difference in the electron charge densities, $\rho_e$, and orientations can greatly affect $\bar{V}_i$, even if the ion positions are identical. To see this, Fig. 2 shows a toy model where a unit cell in a periodic cubic lattice contains a single proton but different $\rho_e$. For case (a) in Fig. 2a, we have a homogeneous $\rho_e$ within a sphere of radius $R = 2$ Å, namely, $\rho_e(r) = 3e/4\pi R^3$ for $r < R$ and $\rho_e(r) = 0$ for $r > R$. Equation (7) yields for the (100) and (120) directions identical $\bar{V}_{(100)} = \bar{V}_{(120)} = -eR^2/10\varepsilon_0 a^3 = -0.072$ eV. This is because the unit cell boundary lines in Fig. 2a do not cut through the electron density in either direction. For case (b) in Fig. 2b, on the other hand, while we also have a homogeneous $\rho_e$, due to the larger radius $R = 4.5$ Å, $\rho_e$ must cross the boundaries. In this case, $\bar{V}_{(100)}$ and $\bar{V}_{(120)}$ are –0.366 and –0.164 eV, respectively. Compared to $\bar{V}_{(100)}$, $\bar{V}_{(120)}$ is reduced in this case because of the overlap of the projected $\rho_e$ along the (120) direction (Fig. 2d,f).

In Ref. [15], a different approach was introduced to identify the ideal vacuum level $\phi_0$ in bulk, namely, the vacuum insertion method, where it was found that $\phi_0$ is equal to the maximum value of the planar average electrostatic potential. It is straightforward to show that the two approaches are identical. For the model cases at hand, Figs. 2c-f depict the average electrostatic potentials with respect to $\phi_0$ (denoted as $\bar{V}_{\phi_0}$ in Fig. 2c) calculated using the method in Ref. [15]. In all cases, a perfect agreement between the quadrupole approach and the maximum value of the planar average potential was found.

By the above discussions, it is clear that eigenvalues associated with the intrinsic band structure of a solid with respect to the true vacuum level, $\varepsilon_{\phi_0}$, should be orientation-dependent due to the non-vanishing bulk quadrupole. For applications where interface, and hence band alignment, is *not* involved, however, it would still be valuable to have an orientation-independent band structure, e.g., with respect to the bulk average potential, $\varepsilon_{\overline{V}}$. The difference between $\varepsilon_{\phi_0}$ and $\varepsilon_{\overline{V}}$ is just $\phi_0 - \overline{V}$. Conceptually, there is no obstacle against decomposing $\varepsilon_{\phi_0}$ into (orientation-independent) $\varepsilon_{\overline{V}}$ + (orientation-dependent) $\phi_0 - \overline{V} = 2\pi \mathbf{n}^{\mathrm{T}} \boldsymbol{Q} \mathbf{n} / \Omega$. To see this, we perform density functional theory (DFT) calculations using Perdew-Burke-Ernzerhof (PBE) exchange-correlation functional [34] with projector augmented wave (PAW) pseudopotentials [35], as implemented in the VASP code [36]. Figure 3 shows the relations between the valence band maximum (VBM), $\overline{V}$, and $\phi_0$ along three different crystallographic orientations of diamond. While the VBM relative to $\overline{V}$, $\varepsilon_{\overline{V}}^{\mathrm{VBM}}$, does not depend on the orientations, the quadrupole term shows a strong orientation dependence as expected.

Determination of the offset between $\phi_0$ and $\overline{V}$ reshapes our understanding of the band offsets at heterojunctions. While $\overline{V}$ has often been regarded as a reference potential when describing the interfacial effects on the band offsets [10,12], our findings indicate that such calculations of the interfacial charge transfer dipole are incorrect as they neglect all of the contribution from the bulk quadrupole. Truncation of the bulk quadrupole leaves a dipole at the surface which is often large but purely reflective of the bulk material properties and has nothing to do with interfacial charge transfer. One the other hand, there is indeed an additional dipole associated with the charge transfer which depends on the detailed chemistry at the interface. This suggests an explanation to the unexpectedly large interface dipoles between quite similar materials [10], as in such cases the dipole is dominated by bulk quadrupole terms instead of the interfacial chemistry.

Our DFT calculation using HSE functionals [37] indeed shows that the bulk quadrupole can significantly contribute to the band offsets. Table 1 shows comparisons between the calculated and measured valence band offsets (VBOs) in some selected traditional semiconductor heterojunctions, Si/GaP, AlP/GaP, and Ge/AlAs. Such heterojunctions are chosen as they possess large bulk quadrupole differences between the constituent materials. The $\overline{V}$- and $\phi_0$-aligned VBOs are calculated by setting the reference energies $\overline{V} = 0$ and $\phi_0 = 0$, respectively. As shown in Table 1, the $\overline{V}$-aligned VBOs are misleading because they attribute the huge differences of VBOs (1.74±0.08 eV) compared to the experiment to interfacial charge-transfer effects. In $\phi_0$-aligned cases, on the other hand, the calculated VBOs are much closer to the experiment, with the differences ranging from 0.04 to 0.35 eV – about an order of magnitude smaller than those obtained from $\bar{V}$-aligned VBOs. Based on $\phi_0$-aligned VBOs, the interfacial charge transfer at traditional semiconductor heterojunctions is small (consistent with the nature of the chemical bonding). The discrepancy between $\bar{V}$- and $\phi_0$-aligned VBOs is exactly the bulk quadrupole effect, which should not be neglected.

This finding is expected to impact the basic understanding in condensed-matter physics and chemistry, in particular in the interpretation of first-principles calculations. As we now have an absolute potential reference between periodic systems, eigenvalue shifts in bulk materials can now be precisely determined and understood. For instance, to improve the band gap in conventional DFT calculations, many-body correction schemes such as HSE and GW [38,39] have been widely used. However, the difficulty has been how to compare the eigenvalues between the different schemes. In general, $\bar{V}$ has been routinely used as the reference potential without justification. A well-defined energy reference $\phi_0$ here enables a precise calculation of the many-body corrections to the eigenvalues of solid as well as the band offsets at heterojuction interfaces [26,40]. This also opens new avenues in interface science, such as to what degree the band offset is determined by the intrinsic quadrupole alignment (for example, as shown in Table 1) and to what degree can it be modified by interfacial impurities or chemistry. Such understanding could significantly advance our ability to design the functionality of interfaces. Furthermore, the *absolute* electrode potentials and redox potentials at electrochemical cells can now be calculated without any external energy reference. This can be done by replacing the conventional energy reference, namely, the local vacuum levels just outside the electrolyte surfaces [41], by $\phi_0$ of the electrolyte. Such procedure eliminates the contribution of the electrolyte surfaces properties to the electrode and redox potentials, which do not contribute in any way to the actual electrochemical processes.

In conclusion, we have introduced an analytic real-space unfolding scheme to uncover the vacuum level $\phi_0$ in infinite solids. We have derived its relationship with the standard average electrostatic potential $\bar{V}$ whose offset with $\phi_0$ is given by an orientation-dependent bulk quadrupole, which can now be unambiguously and accurately calculated. These results provide new insights into the hereto hidden intrinsic physical properties of infinite solids, as well as laying a solid ground for understanding the electronic changes at interfaces with respect to the non-contacting solids (on a par with modern molecular orbital theory for quantum chemistry). Last but not least, we offer a decomposition of the band structure of infinite solids into an orientation-independent term and an orientation-dependent term, and demonstrate the importance of quadrupole contribution to the band alignment at heterojunctions.

**Acknowledgement**

This work was supported by the US DOE Grant No. DESC0002623. The supercomputer time sponsored by NERSC under DOE contract No. DE-AC02-05CH11231 and the CCI at RPI are also acknowledged.

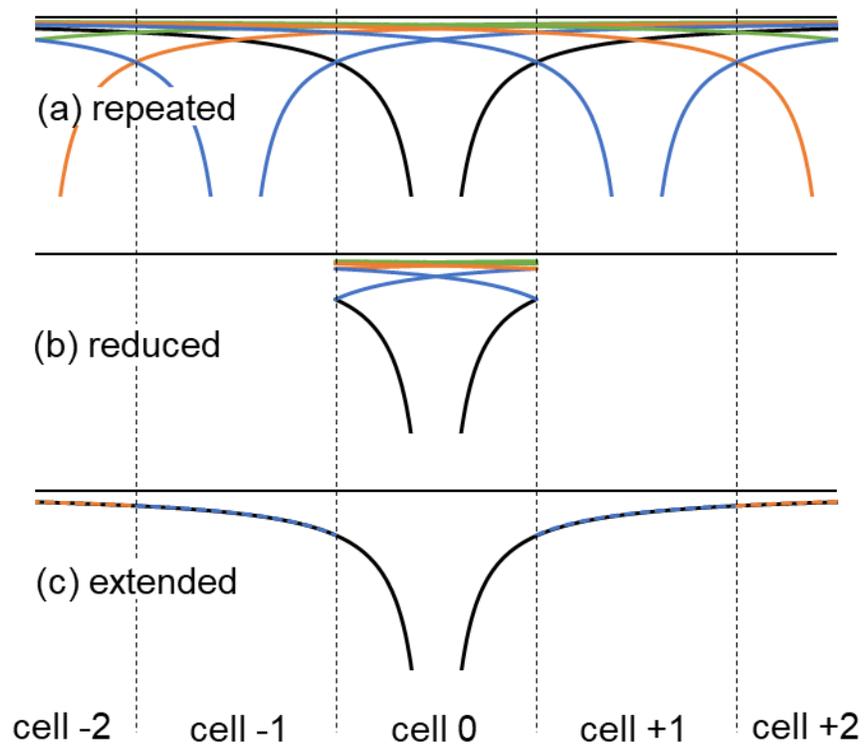

**Fig. 1.** Electrostatic potentials of infinitely large crystals in (a) repeated, (b) reduced, and (c) extended schemes. The potentials due to the 0th (central), ±1th, ±2th, and ±3th cells are represented by the black, blue, orange, and green solid lines, respectively. In (c), the unfolded potential of (b) is shown by a combination of the black solid and dashed colored lines in different cells, which is identical to the electrostatic potential due to the 0th unit cell (i.e., the black solid line in the whole space).

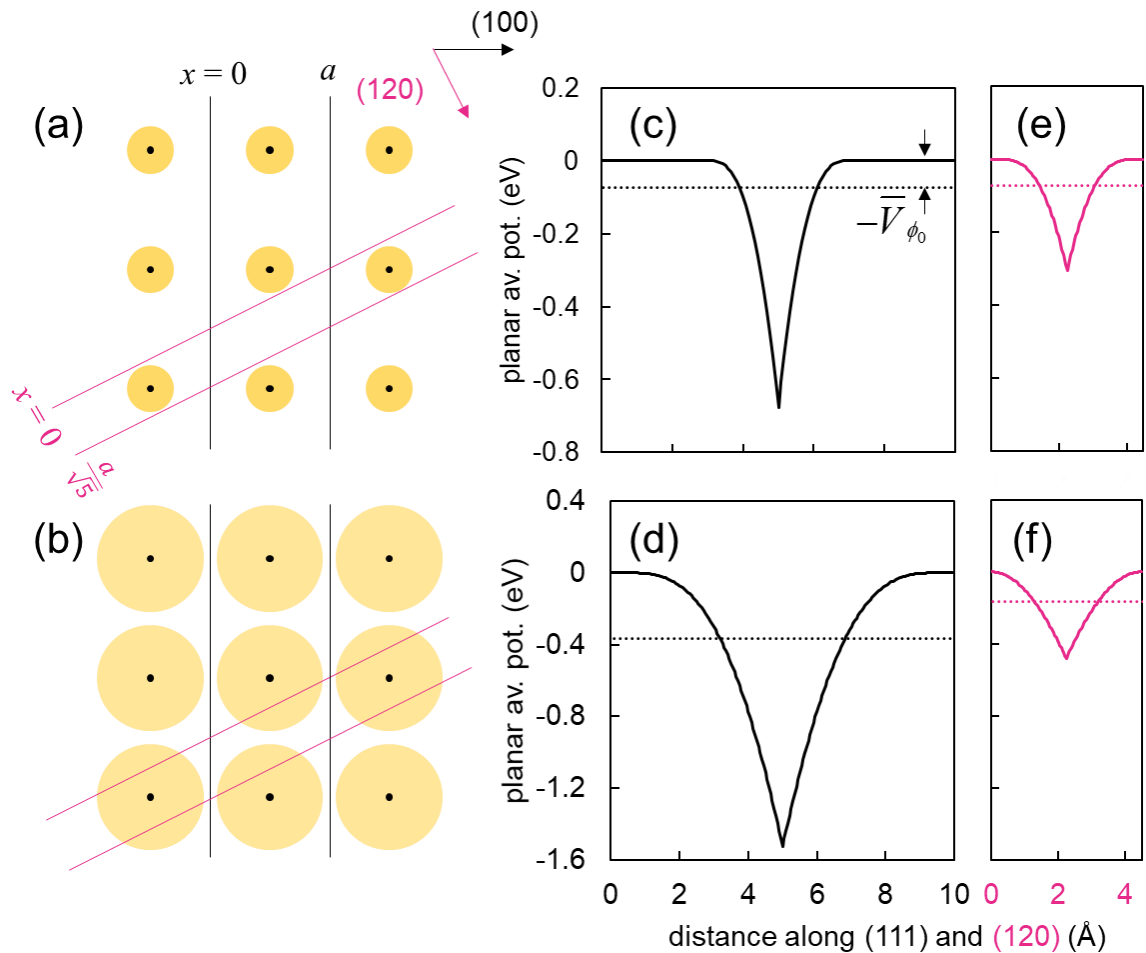

**Fig. 2.** Lattice models (a,b) depicting charge distributions and their corresponding planar average electrostatic potentials (c-f), whose maximum value, equivalent to $\phi_0$, has been set to zero. In the potential plots, dotted lines denote the calculated average electrostatic potential.

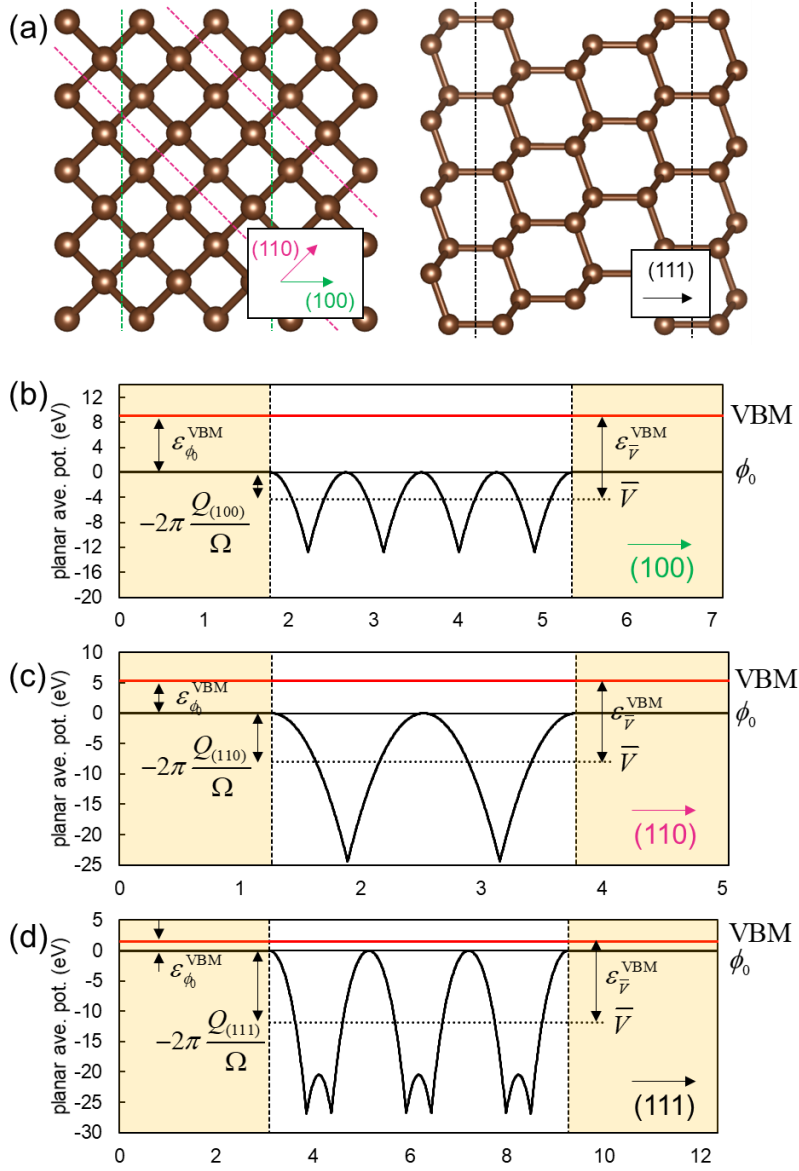

**Fig. 3.** (a) Atomic structures of bulk diamond. (b-d) Planar average electrostatic potentials (PAEP) along (100), (110), and (111) with an ideal vacuum insertion (see the yellow-shaded regions) where the vacuum level $\phi_0$ is set to zero. The PAEP and quadrupole are calculated using the total charge density, i.e., the *core electron density + valence electron density + point ionic charges*. The average potentials and VBM are given by dashed black and red lines, respectively.

**Table 1.** Comparison between $\bar{V}$-aligned, $\phi_0$-aligned, and experimental VBOs. Experimental values are taken from ref. [26] and therein. Units are in eVs.

|  | $\bar{V}$-aligned | $\phi_0$-aligned | Experiment |
|---|---|---|---|
| Si/GaP(110) | –0.97 | 0.84 | 0.80 |
| AlP/GaP(100) | –2.23 | –0.22 | –0.57, –0.43 |
| Ge/AlAs(110) | 2.74 | 1.04 | 0.95, 0.9 |